\documentclass[prl,twocolumn,graphics,floatfix,mathbbm,
superscriptaddress,nofootinbib]{revtex4-2}
\usepackage{amsthm}
\usepackage{amsmath}
\usepackage{latexsym}
\usepackage{amsfonts}
\usepackage{amssymb}
\usepackage{xcolor}
\usepackage{bbm,dsfont}
\usepackage{graphicx}
\usepackage{mathrsfs}
\usepackage{mathbbol}
\usepackage{hyperref}
\usepackage{enumerate}
\usepackage{MnSymbol}
\usepackage{bbding}
\usepackage{float}
\usepackage{soul}
\usepackage{ulem}
\usepackage{braket}

\begin{document}

\title{Metastable discrete time-crystal resonances in a dissipative central spin system}

\author{Albert Cabot}
\affiliation{Institut für Theoretische Physik, Eberhard Karls Universität Tübingen, Auf der Morgenstelle 14, 72076 Tübingen, Germany}
\author{Federico Carollo}
\affiliation{Institut für Theoretische Physik, Eberhard Karls Universität Tübingen, Auf der Morgenstelle 14, 72076 Tübingen, Germany}
\author{Igor Lesanovsky}
\affiliation{Institut für Theoretische Physik, Eberhard Karls Universität Tübingen, Auf der Morgenstelle 14, 72076 Tübingen, Germany}
\affiliation{School of Physics and Astronomy and Centre for the Mathematics and Theoretical Physics of Quantum Non-Equilibrium Systems, The University of Nottingham, Nottingham, NG7 2RD, United Kingdom}

\begin{abstract}
We consider the non-equilibrium behavior of a central spin system where the central spin is periodically reset to its ground state. The quantum mechanical evolution under this effectively dissipative dynamics is described by a discrete-time quantum map. Despite its simplicity this problem shows surprisingly complex dynamical features. In particular, we identify several metastable time-crystal resonances. Here the system does not relax rapidly to a stationary state but undergoes long-lived oscillations with a period that is an integer multiple of the reset period. At these resonances the evolution becomes restricted to a low-dimensional state space within which the system undergoes a periodic motion. Generalizing the theory of metastability in open quantum systems, we develop an effective description for the evolution within this long-lived metastable subspace and show that in the long-time limit a non-equilibrium stationary state is approached. Our study links to timely questions concerning emergent collective behavior in the ``prethermal" stage of a dissipative quantum many-body evolution and may establish an intriguing link to the phenomenon of quantum synchronization.
\end{abstract}

\maketitle

{\it Introduction.---} The interplay of coherent and incoherent processes in interacting driven-dissipative quantum systems can lead to non-equilibrium phases and symmetry breaking \cite{Diehl2008,Diehl2010,Lee2013,Jin2013,Marcuzzi2014,Marcuzzi2016,Buonaiuto2021}, the emergence of long relaxation time scales \cite{Kessler2012,Macieszczak2016,Casteels2017} or dynamical hysteresis \cite{Casteels2016}. Time crystals are an example of a genuine non-equilibrium phase \cite{Wilczek2012,Bruno2013,Watanabe2015}, in which discrete or continuous time-translation symmetry is spontaneously broken. Such phases were initially reported for Hamiltonian systems subject to periodic driving \cite{Else2016,Khemani2016,Yao2017}, and later also found in driven-dissipative scenarios \cite{Iemini2018,Gong2018,Tucker2018,Wang2018}. They have also been reported in a prethermal regime \cite{Else2017,Machado2020,Else2020,Kyprianidis2021}, manifesting in long-lived period-doubling collective oscillations. Emergent long time scales are also common in the dynamics of open quantum systems  \cite{Macieszczak2016,Rose2016,Boite2017,Wolff2020,Cabot2021,Macieszczak2021,Labay2022}. They occur typically in the vicinity of dissipative phase transitions (DPTs), as a finite-size manifestation of the closure of the Liouvillian spectral gap \cite{Minganti2018,Fitzpatrick2017,Fink2018}. %, in close analogy to equilibrium quantum phase transitions \cite{Kessler2012}. 
Similarly, the emergence of multistability and symmetry broken phases can be accompanied by such long-lived dynamical response \cite{Minganti2018}. Metastable dynamics, i.e. ``prestationary" regimes characterized by relaxation into long-lived states \cite{Macieszczak2016}, may also emerge independently of any stationary phase transition, as a purely dynamical phenomenon. This is, e.g., the case in constrained spin models \cite{Lesanovsky2013,Rose2022}. Such long-lived excitations can  impose their frequency to the system, leading to quantum synhronization phenomena  \cite{Giorgi2019,Cabot2019,Tindall2020,Cabot2021}.

\begin{figure}[t]
\includegraphics[width=1\linewidth]{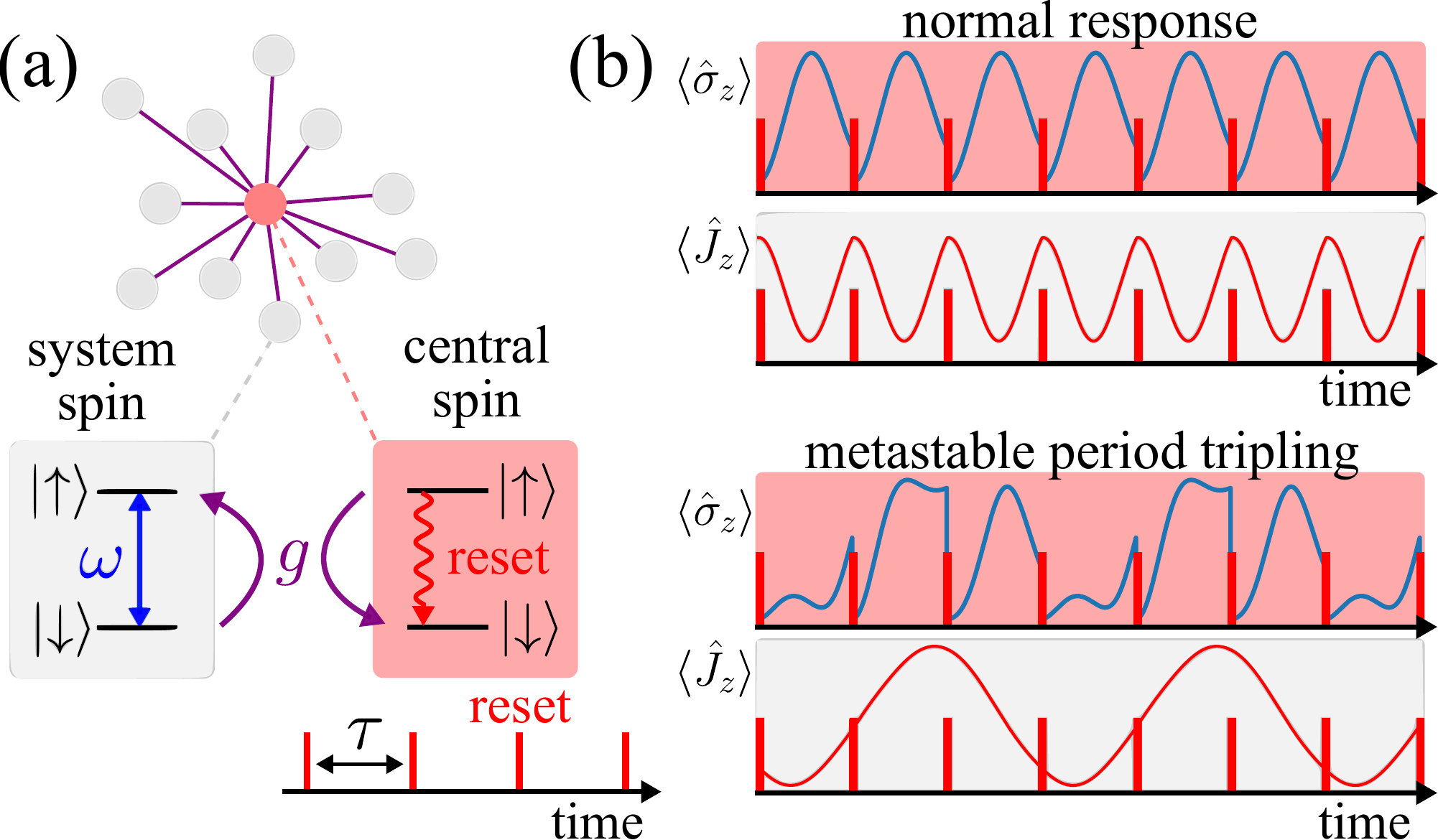}\\
\caption{\textbf{Dissipative central spin model.} (a) $N$ system spins interact with a central spin for a time period $\tau$ after which the latter is reset to the ground state. (b) Long-time dynamics of the magnetization of the central spin, $\langle \hat{\sigma}_z \rangle$, and of the system spins, $\langle \hat{J}_z \rangle$. In the normal response, both the system spins and the central spin approach a steady state that oscillates with the reset period $\tau$ (upper panels). In the metastable time-crystal regime, the spins display long-lived oscillations whose period is a multiple of $\tau$; in the present case $3\tau$ (lower panels). Parameters: $N=30$, $g\tau=0.2$ and $\omega\tau =1.5$ (upper panels) or $\omega\tau=2\pi/3$ (lower panels).}\label{fig_cartoon}
\end{figure}

\begin{figure}[t]
\includegraphics[width=1\linewidth]{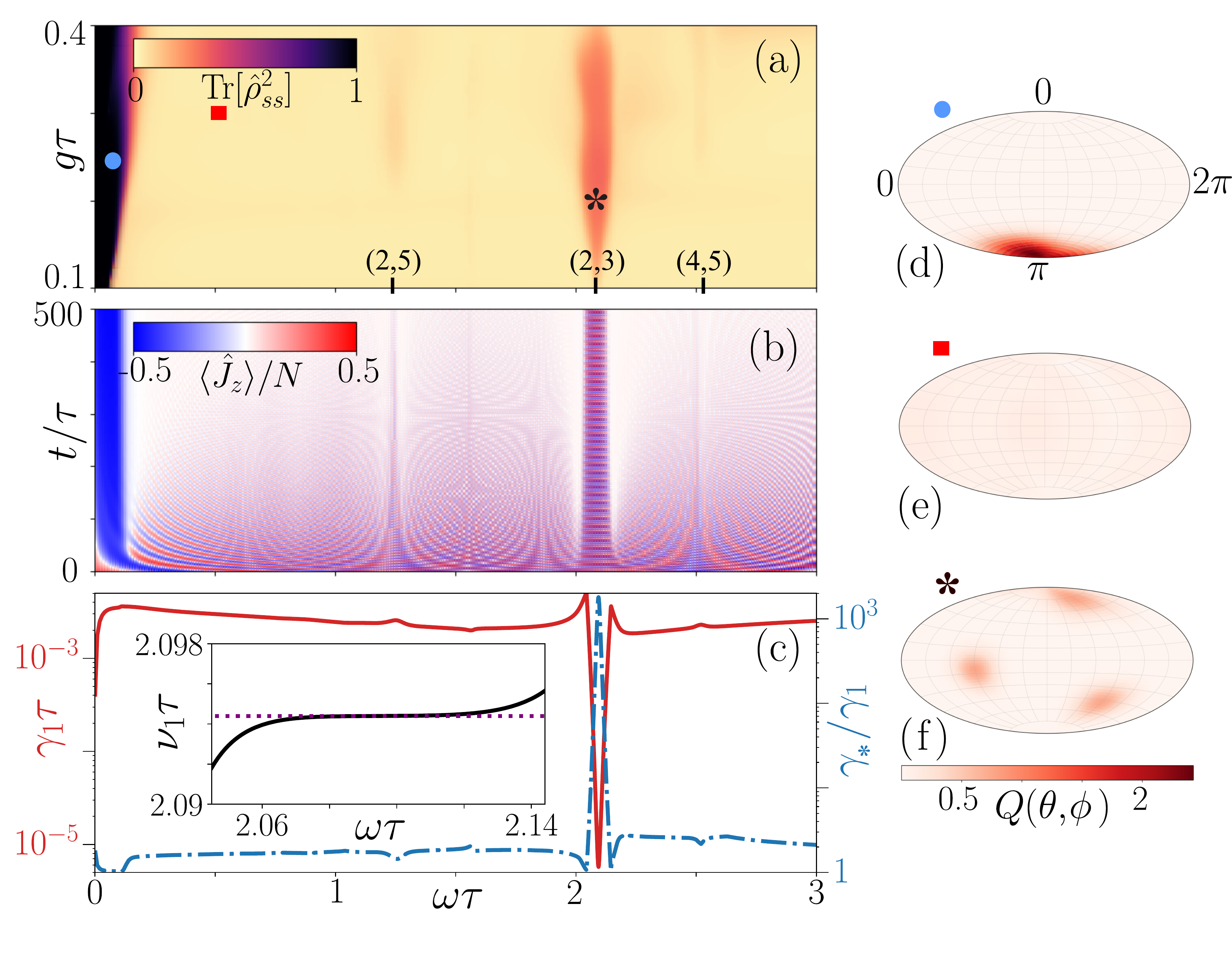}\\
\caption{\textbf{Dynamical regimes and metastable resonances.} (a) Purity of the stationary state as a function of $g$ and $\omega$. Three $(p,q)$ resonances are marked. (b) Initial relaxation dynamics of the system magnetization with initial state $|J,J\rangle$, as a function of $\omega$ for $g\tau=0.2$. (c) Red solid line: decay rate $\gamma_1$ of the leading excitation mode of the map $\mathcal{E}$ [see Eq. (\ref{CM_dynamics})] as a function of $\omega$ at $g\tau=0.2$. Blue dashed-dotted line: ratio of the leading decay rate and the next (different) one, $\gamma_*$ (for the $(p,q)=(2,3)$ resonance this is $\gamma_*=\gamma_3$). Inset: leading frequency, $\nu_1\tau$, of the metastable resonance. The dotted purple line corresponds to $\nu_1\tau=2\pi/3$. (d)-(f) Aitoff projection of the Husimi Q function for $\hat{\rho}_{\mathrm{ss}}$ and for the parameters indicated by the markers in (a):  (d) $g\tau=0.25$, $\omega\tau =0.1$; (e) $g\tau=0.3$, $\omega\tau=0.5$; (f) $g\tau=0.2$, $\omega\tau=2\pi/3$. North to south pole: $\theta=0$ to $\theta=\pi$. West to east: $\phi=0$ to $\phi=2\pi$. In all cases $N=30$.}\label{fig_regimes}
\end{figure}

In this paper, we consider an open quantum dynamics realized by interrupting the coherent evolution of a system with the periodic resetting of some of its degrees of freedom [see Fig. \ref{fig_cartoon} (a)]. By focussing on a central spin system, we show that this dissipative discrete dynamics can give rise to novel non-equilibrium phenomena. In particular, we observe the emergence of metastable resonances, in which the system displays long-lived oscillations, with a period that is locked to a multiple of the reset period [see Figs. \ref{fig_cartoon}(b) and \ref{fig_regimes}], and in which ``heating'' towards an infinite-temperature state is avoided. To describe the dynamics at these resonances, we develop an effective theory in which the time-translation symmetry breaking is decoupled from an emergent classical non-equilibrium dynamics accounting for the (slow) eventual decay towards the stationary state. Our results demonstrate the emergence of non-equilibrium behavior that is substantially different from conventional phase transitions and ``prethermal" time-crystalline phases. The observed behavior is metastable in the sense that its manifestation requires large system sizes, and is seemingly not related to a (standard) non-equilibrium phase transition. Upon varying the system size, the metastable resonances can become weaker or stronger, and do not smoothly approach a well-determined thermodynamic limit.

{\it Dissipative central spin model.---} We consider a central spin model in which $N$ system spins interact with a central one. Such models are relevant in the description of hyperfine interactions between quantum dots \cite{Urbaszek2013} or nitrogen-vacancy centers in diamond \cite{Schwartz2018} and their  environment. Moreover, they are known to display interesting dynamical phenomena both in closed and driven-dissipative scenarios \cite{Bortz2007,Villazon2020a,Villazon2020b,Kessler2010,Kessler2012}. %as integrable regimes \cite{Bortz2007,Villazon2020a}, persistent dark states \cite{Villazon2020b}, superradiance \cite{Kessler2010} or DPTs \cite{Kessler2012}. 
Our starting point is an $XX$-Hamiltonian ($\hbar=1$) where all system spins are resonantly driven with a Rabi frequency $\omega$ and coupled with the same strength $g$ to the central spin [see Fig. \ref{fig_cartoon}(a)]: 
\begin{equation}\label{Ham}
\hat{H}=\omega \hat{J}_x+g\big(\hat{J}_+\hat{\sigma}_- + \hat{J}_-\hat{\sigma}_+\big).    
\end{equation}
Here $\hat{J}_\alpha=\frac{1}{2}\sum_{j=1}^N \hat{\sigma}_\alpha^{(j)}$ and $\hat{J}_\pm=\hat{J}_x\pm i\hat{J}_y$ are collective spin operators and raising/lowering operators, respectively, representing the ensemble of system spins. They are constructed from the Pauli matrices $\hat{\sigma}_\alpha$ ($\alpha=x,y,z$). The central spin is represented through the raising/lowering operators $\hat{\sigma}_\pm=(\hat{\sigma}_x\pm i\hat{\sigma}_y)/2$. %Hence, Hamiltonian (\ref{Ham}) models the system in a rotating frame assuming that all spins have the same frequency. 
The dissipative dynamics emerges from periodically (period  $\tau$)  resetting to the ground state ($\ket{\downarrow}\bra{\downarrow}_c$) the central spin, as illustrated in Fig. \ref{fig_cartoon} (a). The reduced density matrix $\hat{\rho}$ for the collective spin at multiples of $\tau$ is then given by \cite{CMintroduction}
\begin{equation}\label{CM_dynamics}
\hat{\rho}_{n+1}=\mathcal{E}\hat{\rho}_n=\text{Tr}_c\big[\hat{U} \hat{\rho}_n\otimes \mid\downarrow\rangle\langle \downarrow\mid_c \hat{U}^\dagger \big].    
\end{equation}
Here $\hat{U}=e^{-i\hat{H}\tau}$ and $\hat{\rho}_n$ is the short hand notation for $\hat{\rho}(n\tau)$. The map $\mathcal{E}$ is a trace-preserving and completely positive quantum map. This kind of discrete quantum dynamics also occurs in so-called collision models \cite{CMintroduction}, which in the short interaction time limit provide a dynamics that is equivalent to Lindblad master equation. 

{\it Dynamical regimes and stationary purity. ---}  The action of the map $\mathcal{E}$ is conveniently studied in terms of its right and left eigenmatrices, i.e., $\hat{R}_j$ and $\hat{L}_j$, and the corresponding eigenvalues $\lambda_j$ \cite{Macieszczak2016}:
\begin{equation}\label{state_deco}
\hat{\rho}_n=\mathcal{E}^n  \hat{\rho}_0=\hat{\rho}_{\mathrm{ss}}+\sum_{j\geq 1}\text{Tr}[\hat{L}_j\hat{\rho}_0]\hat{R}_j \lambda_j^n.  
\end{equation}
The eigenvalues satisfy $|\lambda_j|\leq1$, and we arrange them in order of decreasing absolute value $|\lambda_0|\geq|\lambda_1|\geq |\lambda_2|\geq \dots$. Those with unit absolute value correspond to non-decaying modes, while there is at least one stationary state that we denote as $\hat{\rho}_{\mathrm{ss}}=\hat{R}_0/\text{Tr}[\hat{R}_0]$ with $\lambda_0=1$. It is also useful to define the frequencies and decay rates of the system as: $\gamma_j \tau=-\text{ln}[|\lambda_j|]$, $\nu_j\tau=\text{arg}[\lambda_j]$, in analogy to the Liouvillian formalism \cite{Minganti2018,Macieszczak2016}. Additionally, we will make use of spin coherent states, which provide the basis for a phase space representation of the system state and operators. This representation provides important insight on the different dynamical regimes of the model. For a spin $J$, coherent states are defined as $|\theta,\phi\rangle=\exp(-i\phi \hat{J}_z)\exp(-i\theta \hat{J}_y)|J,J\rangle$, where $\theta \in [0,\pi]$ and $\phi\in[0,2\pi]$ define the polar and azimutal angle respectively \cite{Roulet2018}. Following \cite{Roulet2018}, we will make use of the spin analogous of the Husimi Q function, defined for a spin operator $\hat{O}$ as: $
Q(\theta,\phi)=\mathcal{J}\langle \theta, \phi|\hat{O}|\theta,\phi\rangle$,  where  $\mathcal{J}=(2J+1)/4\pi$ is a normalization constant.

We find that for the study of the stationary state $\hat{\rho}_{\mathrm{ss}}$ the purity is actually a good order parameter, noting that the magnetization, which appears to be the natural order parameter, is not sensitive to all dynamical regimes. The stationary purity is shown in Fig. \ref{fig_regimes} (a) as a function of the Rabi frequency and interaction strength. For small $\omega\tau$, $\hat{\rho}_{\mathrm{ss}}$ is almost pure, which follows from the interaction term dominating over the coherent driving, the former enforcing a stationary state close to $|J,-J\rangle$ [Fig. \ref{fig_regimes} (d)]. In contrast, increasing $\omega\tau$ above a certain threshold makes the Rabi term to dominate over the interaction one, changing qualitatively the stationary state. In this region, $\hat{\rho}_{\mathrm{ss}}$ is a highly entropic mixed state, close to the infinite temperature state, and thus spreading out (quasi) uniformly over the entire phase space [Fig. \ref{fig_regimes} (e)].  For even larger $\omega\tau$, the interplay of coherent dynamics and periodic interruptions gives rise to yet another kind of dynamics: for $\omega\tau$ close to certain fractions of $\pi$, i.e. $p\pi/q$, purity islands emerge in which  $\text{Tr}[\hat{\rho}_{\mathrm{ss}}^2]\sim1/q$. In panel (a) these $(p,q)$ resonances can be observed around $2\pi/5$, $2\pi/3$ and $4\pi/5$, the most prominent one being the one with $q=3$. Near these {\it metastable resonances} the stationary state, rather than being almost fully mixed, is actually a mixture of $q$ almost disjoint and highly pure states. This is illustrated in Fig. \ref{fig_regimes} (f) for $q=3$ and in the supplemental material \cite{SM} for $q=5$. In order to get an impression of the dynamics, we show in Fig. \ref{fig_regimes} (b) the time evolution of the system spin magnetization $\langle \hat{J}_z \rangle$ for $g\tau=0.2$. Here one finds that the purity islands labeled by $(p,q)$ in panel (a) indeed correspond to metastable states, which display long-lived oscillations with a period that is approximately given by $T\approx q\tau$. This behavior is most evident around $(p,q)=(2,3)$, although it can also be recognized near the other resonances [$(2,5)$ and $(4,5)$]. 
In the following, we focus on the case $(p,q)=(2,3)$, as the most prominent resonance in Fig. \ref{fig_regimes}.

\begin{figure*}
\includegraphics[width=1\linewidth]{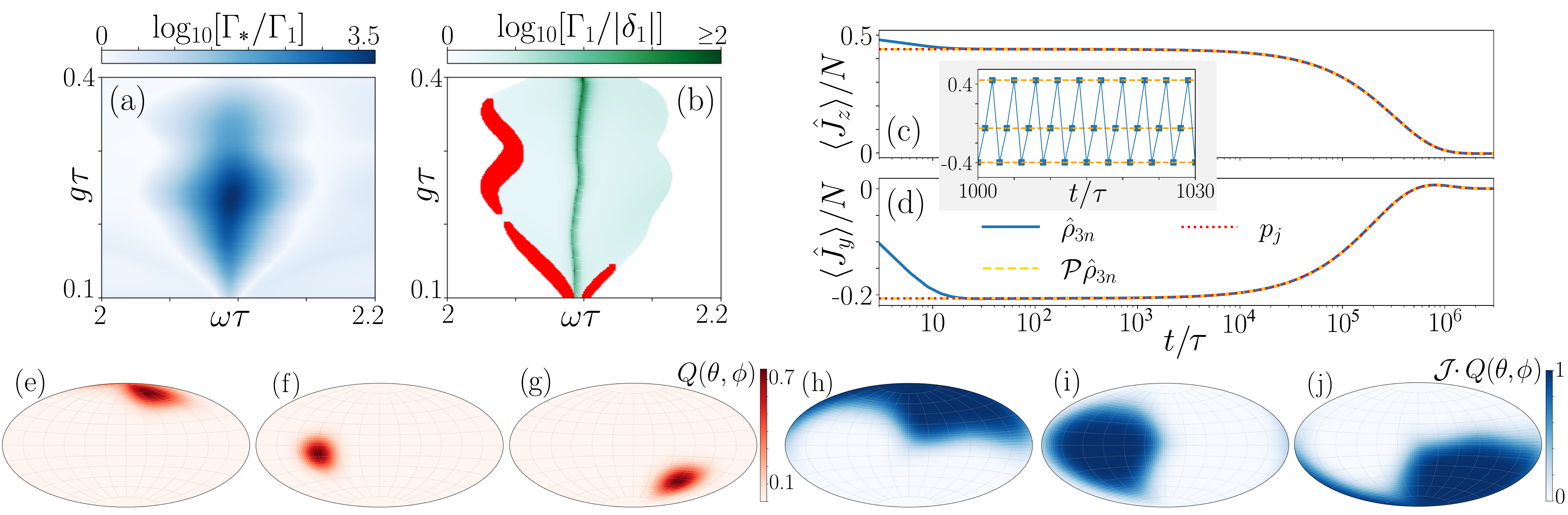}\\
\caption{{\bf Metastable dynamics.} (a) Ratio of the leading decay rate, $\Gamma_1$, and the next (different) one, $\Gamma_*$, for the period-tripled stroboscopic map $\Lambda$. (b) Ratio of the leading decay rate and its corresponding detuning, $\delta_1$, in the region in which the leading mode is complex. The regions in which $\Gamma_1/|\delta_1|<\sqrt{3}$ are colored in red. (c) and (d) Stroboscopic  dynamics of the z- and y-component in the metastable regime considering the exact time evolution $\hat{\rho}_{3n}$ (blue solid line), the projection on the metastable manifold $\mathcal{P}\hat{\rho}_{3n}$ (yellow dashed line) and the classical stochastic model of Eq. (\ref{classical_process}) (red dotted line). Initial condition $|J,J\rangle$, $\omega\tau=2\pi/3$ and $g\tau=0.25$. Gray inset: the blue squares show the exact value of $\langle \hat{J}_z (n\tau)\rangle/N$, while the orange-dashed lines show the values obtained from Eq. (\ref{period3_MM}). Aitoff projection of the Husimi Q function for: (e) $\hat{\mu}_{1}/3$; (f) $\hat{\mu}_{2}/3$ (g) $\hat{\mu}_{3}/3$; (h) $\hat{P}_1$; (i) $\hat{P}_2$; (j) $\hat{P}_3$. Parameters: $N=30$, $\omega\tau=2\pi/3$, $g\tau=0.25$. }\label{fig_meta}
\end{figure*}

{\it Metastable period-locking resonances.---} The observed long-lived oscillations can be characterized studying the leading eigenvalues of $\mathcal{E}$. In Fig. \ref{fig_regimes} (c) we plot the leading decay rate of the system $\gamma_1\tau$ (red solid lines), observing that for the resonance around $\omega\tau=2\pi/3$ this becomes several orders of magnitude smaller than anywhere else. At this resonance the eigenvalue $\lambda_1$ is complex and its corresponding frequency is plotted in the inset. This reveals a {\it frequency locking} to $\nu_1\tau\approx 2\pi/3$ across the entire $(2,3)$ resonance, a behavior that is reminiscent of the synchronization phenomenon of frequency entrainment. Here, the dominant frequency of a system also locks to a given one in a whole dynamical regime, which has been observed both in classical \cite{PikovskyBook} and quantum systems in \cite{Walter2014,Sonar2018,Cabot2021}. In Fig. \ref{fig_regimes} (c) we also plot the ratio of the dominant decay rate with the next (different) leading one (blue dotted-line) \cite{footnote3}. This ratio increases by several orders of magnitude at the $(2,3)$ resonance, indicating the emergence of a huge separation of time scales: the long-time dynamics is thus dominated only by the two complex conjugated modes with frequency $|\nu_1|\tau \approx 2\pi/3$.  This separation of time scales is characteristic of the emergence of metastability in open quantum systems \cite{Macieszczak2016}. Moreover, we find it to be present in the whole purity island, as shown in Fig. \ref{fig_meta} (a), giving rise to what we term as {\it metastable period-locking resonances}. Similar results can be found for resonances with higher $q$ and different system sizes, in which more than two long-lived modes can be involved \cite{SM}.

This huge separation of time scales allows the metastable dynamics in the $(2,3)$ resonance to be approximated by \cite{Macieszczak2016}: 
\begin{equation}\label{long_time}
\hat{\rho}_n\approx\hat{\rho}_{\mathrm{ss}}+(c_1\hat{R}_1e^{i\nu_1n\tau}+\mathrm{H.c})e^{-\gamma_1 n\tau}=\mathcal{P}\hat{\rho}_n,    
\end{equation}
with $c_1=\text{Tr}[\hat{L}_1\hat{\rho}_0]$. The stationary state and these two longest-lived modes define the metastable manifold (MM), denoted by $\mathcal{P}\hat{\rho}_n$, to which the state of the system rapidly converges on a time scale given by $\gamma_{3}^{-1}$. Crucially, while $\gamma_1\tau\ll 1$, $\nu_1\tau$ is of order one. Therefore, in order to unveil the structure of the MM, it is more convenient to consider the {\it period-tripled stroboscopic map} $\Lambda=\mathcal{E}^3$, which displays the same eigenmatrices but with eigenvalues $\delta_j=3\nu_j-2\pi$ and $\Gamma_j=3\gamma_j$. Thus, by switching to $\Lambda$, we  have $\Gamma_1\tau \sim \delta_1\tau \ll 1$, as can be readily appreciated in Fig. \ref{fig_meta} (b). The metastable dynamics in this stroboscopic picture is exemplified in Fig. \ref{fig_meta} (c) and (d), in which its multistep character is evident: the system rapidly relaxes to the MM, in which the state appears to settle to $\hat{\rho}_{3n}\approx \mathcal{P}\hat{\rho}_0$ displaying a {\it metastable plateau} for intermediate times $\Gamma_3^{-1} \ll t\ll \Gamma_1^{-1}, |\delta_1^{-1}|$, until eventually reaching the true stationary state, described by Eq. (\ref{long_time}).

{\it Metastable symmetry broken states. ---} The smallness of both $\Gamma_1$ and $\delta_1$ shows that the metastable resonance manifests in the map $\Lambda$ as a (quasi-)closure of the spectral gap, similarly to what is observed for finite sizes with the Liouvillian gap in DPTs \cite{Kessler2012,Minganti2018}. As a consequence, the structure of the MM is analogous to that of the emerging stationary manifold for DPTs, analyzed in Ref. \cite{Minganti2018}. In fact,  to a good approximation, the stationary state and the Hermitian partners  of the leading eigenmodes \cite{footnote4} decompose in terms of the same three extremal metastable states (EMSs) \cite{Macieszczak2016}, that allow us to write any state in the MM as a convex combination of them [see Eq. (\ref{initial_mixture})]. These EMSs are denoted by $\hat{\mu}_{1,2,3}$ and given by:
\begin{equation}\label{ss_3lobes}
\hat{\rho}_{\mathrm{ss}}\approx (\hat{\mu}_1+\hat{\mu}_2+\hat{\mu}_3)/3,
\end{equation}
\begin{equation}\label{lobes_decomposition}
\hat{R}_\mathrm{A}\approx c_\mathrm{A}(\hat{\mu}_1-\hat{\mu}_2/2-\hat{\mu}_3/2),\quad \hat{R}_\mathrm{B}=c_\mathrm{B} (\hat{\mu}_3-\hat{\mu}_2).   
\end{equation}
Here $c_{\mathrm{A,B}}$ are real constants corresponding to the sum of positive eigenvalues of $\hat{R}_{\mathrm{A,B}}$,  and in the whole metastable resonance their ratio is well approximated  by $c_\mathrm{A}/c_\mathrm{B}\approx 2/\sqrt{3}$ \cite{SM}. In Fig. \ref{fig_meta} (e)-(g) we show these EMS, finding that they correspond to the different lobes making up the stationary state. The accuracy of these approximations is characterized in detail in \cite{SM}. %, in which corrections are shown to be smaller as the separation of time scales increases. 
Moreover, we observe the EMSs to be almost disjoint, as they are tightly focused in non-overlapping phase space regions. This can be better understood considering their left partners: $
\hat{P}_1=\mathbb{1}/3+ c_\mathrm{A}\hat{L}_\mathrm{A}$, $ \hat{P}_2=\mathbb{1}/3-c_\mathrm{A}\hat{L}_\mathrm{A}/2-c_B\hat{L}_\mathrm{B}$, $ \hat{P}_3=\mathbb{1}/3-c_\mathrm{A}\hat{L}_\mathrm{A}/2+c_\mathrm{B}\hat{L}_\mathrm{B}$. These Hermitian operators satisfy the following properties \cite{SM}: (i) $\hat{P}_1+\hat{P}_2+\hat{P}_3=\mathbb{1}$;  (ii) they are to a  good approximation positive; (iii) they satisfy to a good approximation $\text{Tr}[\hat{P}_j\hat{\mu}_k]=\delta_{jk}$. Accordingly, we can rewrite the projection of $\hat{\rho}_0$ in the MM as a probabilistic mixture of $\hat{\mu}_{1,2,3}$:
\begin{equation}\label{initial_mixture}
\mathcal{P}\hat{\rho}_0\approx p^0_1 \hat{\mu}_1+   p^0_2 \hat{\mu}_2+ p^0_3 \hat{\mu}_3, 
\end{equation}
with $p_j^0=\text{Tr}[\hat{P}_j\hat{\rho}_0]$ which can be approximately regarded as classical probabilities. Inspecting the Husimi representation of $\hat{P}_j$, we find them to partition the phase space in three almost disjoint regions [Fig. \ref{fig_meta} (h)-(j)]. These different regions correspond to the basin of attraction of each of the EMSs; any initial state contained in them will rapidly converge to the corresponding EMS and remain trapped in it for the long intermediate time scale.

The action of $\mathcal{E}$ on the EMSs  unveils one of their most interesting features: they break (approximately) the discrete time-translation symmetry imposed by the periodic resetting of the central spin. Indeed, they are connected by $\mathcal{E}$ forming a period-tripled cyclic evolution:
\begin{equation}\label{period3_dynamics}
\mathcal{E}\hat{\mu}_1\approx \hat{\mu}_2,\quad \mathcal{E}\hat{\mu}_2\approx \hat{\mu}_3,\quad 
\mathcal{E}\hat{\mu}_3\approx \hat{\mu}_1.
\end{equation}
These relations can be  derived using  Eqs. (\ref{ss_3lobes})-(\ref{lobes_decomposition}), and making the approximation $c_\mathrm{A}/c_\mathrm{B}\approx2/\sqrt{3}$, $\Gamma_1=\delta_1\approx 0$, valid in the metastable resonance and for times $t \ll \Gamma_1^{-1} \sim |\delta_1|^{-1}$ (i.e. in the metastable plateau) \cite{SM}. Remarkably, the structure of Eq. (\ref{period3_dynamics}) is analogous to that of conventional symmetry breaking DPTs \cite{Minganti2018} replacing the superoperator describing the symmetry, as e.g. parity, by $\mathcal{E}$ [see \cite{SM} for an example with $(p,q)=(4,5)$]. Finally, combining Eqs. (\ref{initial_mixture}) and (\ref{period3_dynamics}), we see that in the metastable plateau the period-tripled dynamics is approximated by: 
\begin{equation}\label{period3_MM}
\hat{\rho}_{3n+j}\approx p_1^0\hat{\mu}_{1+j}+p_2^0\hat{\mu}_{2+j}+p_3^0\hat{\mu}_{3+j},    
\end{equation}
where $j=0,1,2$ and the index of the metastable states follows periodic boundary conditions, i.e. $\hat{\mu}_4=\hat{\mu}_1$. In the inset of Fig. \ref{fig_meta} (c) we consider the exact magnetization dynamics (blue squares) in the plateau and compare it with the three values predicted by this approximation (orange dashed lines), finding excellent agreement.

{\it Effective non-equilibrium  classical relaxation.---}  The   period-tripled oscillation eventually fades away due to the slow residual dynamics associated with the small but non-vanishing values of $\Gamma_1$ and $\delta_1$. By rewriting  Eq. (\ref{long_time}) in terms of the EMSs, we find that this final relaxation (in the stroboscopic picture) follows a classical stochastic process, i.e. not only $\mathcal{P}\hat{\rho}_0$ can be written as a classical probabilistic mixture of the EMSs, but also $\mathcal{P}\hat{\rho}_{3n}$. Hence, we can promote Eq. (\ref{initial_mixture}) to $\mathcal{P}\hat{\rho}_{3n}\approx \sum_{j=1}^3 p_j(n \tau)\hat{\mu}_j$, where the probabilities $p_j(t)$ obey \cite{SM}:
\begin{equation}\label{classical_process} 
\frac{d}{dt}p_j =-\frac{2\Gamma_1}{3}p_j+\bigg(\frac{\Gamma_1}{3}-\frac{\delta_1}{\sqrt{3}} \bigg)p_{j+1}+ \bigg(\frac{\Gamma_1}{3}+\frac{\delta_1}{\sqrt{3}} \bigg)p_{j-1},  
\end{equation}
with initial condition $p_j^0$ and where the index $j$ in (\ref{classical_process}) follows periodic boundary conditions. For the process (\ref{classical_process}) to be physical we need $\Gamma_1/|\delta_1|\geq\sqrt{3}$, which is satisfied in most of the metastable regime, except for the boundary regions indicated in red in Fig. \ref{fig_meta} (b). In Fig. \ref{fig_meta} (c) and (d) we exemplify these dynamics (red-dotted lines), finding excellent agreement with the exact ones both in the plateau  and in the final decay. Inspection of Eq. (\ref{classical_process}) reveals the stationary state to be $p_1=p_2=p_3=1/3$ as expected from Eq. (\ref{state_deco}). Interestingly, we find that  stationary probability currents, given by $J_{j,j+1}=2\delta_1/(3\sqrt{2})$ \cite{SM}, are generally present. This indicates the non-equilibrium nature of the final relaxation process of the metastable time-translation symmetry broken states, that contrasts with what found for other quantum systems effectively governed in the long-time by infinite-temperature classical equilibrium process \cite{Macieszczak2016,Rose2016,Cabot2021}.

{\it Discussion and conclusions.---} We have reported on an intrinsically metastable counterpart of discrete time crystals, emerging in a dissipative spin model that may be regarded as the discrete-time generalization of the boundary time crystal of Ref. \cite{Iemini2018}. Similarly to prethermal time-crystals  \cite{Else2017,Machado2020,Else2020,Kyprianidis2021}, these oscillations emerge in a prestationary regime, while their lifetime surpasses any intrinsic timescale of the model by orders of magnitude. Compared to other many-body ergodicity breaking dynamics, as due to dynamical symmetries \cite{Buca2019} or quantum scars \cite{Serbyn2021}, the reported dynamics is largely independent on the initial conditions, as the EMSs act as effective attractors with a combined basin of attraction that spans all phase space. A further peculiarity is their non-trivial dependence on system size, as the largest lifetimes  are attained for intermediate sizes resulting in a non-monotonic behavior as the thermodynamic limit is approached \cite{SM}. Nevertheless, the fact that this spectral gap does not actually close does not preclude the emergence of a robust MM with a structure analogous to the stationary one found in symmetry breaking DPTs \cite{Minganti2018}. This points to a connection between time-translation symmetry breaking and other types of spontaneous symmetry breaking in driven-dissipative systems. %In conclusion, our results illustrate that non-unitary interruptions of coherent dynamics provide an additional route to engineer novel non-equilibrium phenomena in quantum systems.
\\

\acknowledgements \textbf{Acknowledgements.---} We acknowledge support from the “Wissenschaftler R\"uckkehrprogramm GSO/CZS” of the Carl-Zeiss-Stiftung and the German Scholars Organization e.V., as well as from the Baden-W\"urttemberg Stiftung through Project No.~BWST\_ISF2019-23. We also acknowledge funding from the Deutsche Forschungsgemeinsschaft (DFG, German Research Foundation) under Projects No. 435696605 and 449905436, as well as through the Research Unit FOR 5413/1, Grant No. 465199066.

\setcounter{equation}{0}
\setcounter{figure}{0}
\setcounter{table}{0}
\makeatletter
\renewcommand{\theequation}{S\arabic{equation}}
\renewcommand{\thefigure}{S\arabic{figure}}

\makeatletter
\renewcommand{\theequation}{S\arabic{equation}}
\renewcommand{\thefigure}{S\arabic{figure}}

\onecolumngrid
\newpage

\setcounter{page}{1}

\begin{center}
{\Large SUPPLEMENTAL MATERIAL}
\end{center}
\begin{center}
\vspace{0.8cm}
{\Large Metastable discrete time-crystal resonances in a dissipative central spin system}
\end{center}
\begin{center}
Albert Cabot$^{1}$, Federico Carollo$^{1}$, Igor Lesanovsky$^{1,2}$
\end{center}
\begin{center}
$^1${\it Institut f\"ur Theoretische Physik, Universit\"at T\"ubingen,}\\
{\it Auf der Morgenstelle 14, 72076 T\"ubingen, Germany}\\
$^2${\it School of Physics and Astronomy and Centre for the Mathematics and Theoretical Physics of Quantum Non-Equilibrium Systems, The University of Nottingham, Nottingham, NG7 2RD, United Kingdom}
\end{center}

\section{Period-tripled metastable oscillations}

\subsection{Long-time dynamics}

We begin this section by writing down the long time approximation for the dynamics in the period-tripled resonance:
\begin{equation}\label{longtime_approx}
\hat{\rho}_n\approx\hat{\rho}_{\mathrm{ss}}+\text{Tr}[\hat{L}_1\hat{\rho}_0]\hat{R}_1 e^{i\nu_1 n\tau -\gamma_1n\tau}+  \text{Tr}[\hat{L}^\dagger_1\hat{\rho}_0]\hat{R}^\dagger_1 e^{-i\nu_1 n\tau -\gamma_1n\tau}=\mathcal{P}\hat{\rho}_n. 
\end{equation}
Notice that we stick to the following criteria for the definition of the eigenvalues and eigenmatrices: $\nu_1>0$ and $\text{Tr}[\hat{L}_1 \hat{R}_1]=1$. We want to rewrite Eq. (\ref{longtime_approx}) in terms of the following Hermitian combinations of the long-lived eigenmodes:
\begin{equation}\label{Hermitian_partners}
\hat{R}_\mathrm{A}=\frac{\hat{R}_1+\hat{R}_1^\dagger}{2}, \quad \hat{R}_\mathrm{B}=\frac{\hat{R}_1-\hat{R}_1^\dagger}{2i},\quad
\hat{L}_\mathrm{A}=\hat{L}_1+\hat{L}_1^\dagger,\quad 
\hat{L}_\mathrm{B}=i(\hat{L}_1-\hat{L}_1^\dagger),
\end{equation}
which  satisfy $\text{Tr}[\hat{L}_j \hat{R}_k]=\delta_{jk}$  with $j,k=\mathrm{A,B}$. We then obtain:
\begin{equation}
\mathcal{P}\hat{\rho}_n=\hat{\rho}_{\mathrm{ss}}+[A\cos(\nu_1 n\tau)+B\sin (\nu_1  n\tau) ]\hat{R}_\mathrm{A} e^{-\gamma_1  n\tau} +[B\cos(\nu_1 n\tau)-A\sin (\nu_1  n\tau) ]\hat{R}_\mathrm{B} e^{-\gamma_1  n\tau} ,
\end{equation}
where $A=\text{Tr}[\hat{L}_\mathrm{A}\hat{\rho}_0]$ and $B=\text{Tr}[\hat{L}_\mathrm{B}\hat{\rho}_0]$. The advantage of $\hat{R}_{\mathrm{A,B}}$ comes from the fact that they can be easily decomposed in terms of physical states as they are both Hermitian and traceless. Therefore, they are bound to satisfy:
\begin{equation}
\hat{R}_\mathrm{A}=\sum_{j=1}^{2J+1} a_j|A_j\rangle\langle A_j|,\quad   \hat{R}_\mathrm{B}=\sum_{j=1}^{2J+1} b_j|B_j\rangle\langle B_j|,\quad \sum_{j=1}^{2J+1}a_j=0,\quad \sum_{j=1}^{2J+1} b_j=0,  
\end{equation}
where $2J+1$ is the dimension of the Hilbert space in which the collective spin $J$ resides. It is also useful to write the time evolution every three steps, i.e. in the stroboscopic period-tripled picture:
\begin{equation}\label{stroboscopic_dynamics}
\mathcal{P}\hat{\rho}_{3n}=\hat{\rho}_{\mathrm{ss}}+[A\cos(\delta_1 n\tau)+B\sin (\delta_1  n\tau) ]\hat{R}_\mathrm{A} e^{-\Gamma_1  n\tau} +[B\cos(\delta_1 n\tau)-A\sin (\delta_1  n\tau) ]\hat{R}_\mathrm{B} e^{-\Gamma_1  n\tau} ,
\end{equation}
where $\delta_1=3\nu_1-2\pi$ and $\Gamma_1=3\gamma_1$.

\subsection{Formal decomposition in stationary period-tripled lobes}

In this subsection we will suppose  that $\hat{R}_1$ and $\hat{R}^\dagger_1$ actually become eigenmodes with unit eigenvalue of the period-tripled map and explore the consequences of this. More precisely, we are assuming that $\Lambda \hat{R}_1=\hat{R}_1$, $\Lambda \hat{R}^\dagger_1=\hat{R}^\dagger_1$ and thus $\Lambda \hat{R}_{\mathrm{A,B}}=\hat{R}_{\mathrm{A,B}}$ (recall $\Lambda=\mathcal{E}^3$). Then, $\hat{\rho}_{\mathrm{ss}}$ and $\hat{R}_{\mathrm{A,B}}$ all belong to the now {\it degenerate} stationary subpsace of $\Lambda$. Notice that while $\hat{\rho}_{\mathrm{ss}}$ and $\hat{R}_{\mathrm{A,B}}$ need not to be orthogonal, they are required to be linearly independent (in the vectorized representation). The other constrains to be satisfied are their unit or null trace, their Hermiticity, and the positvity of the stationary state. A possible way to satisfy these constrains is that they are all different linear combinations of a set of three  stationary states. This crucial idea is one of the core results of Ref. \cite{Minganti2018}. Indeed, based on our numerical observations, we propose the following decomposition:
\begin{equation}\label{deco_modes}
\hat{R}_\mathrm{A}=c_\mathrm{A}(\hat{\mu}_1-\hat{\mu}_2/2-\hat{\mu}_3/2),\quad  \hat{R}_\mathrm{B}=c_\mathrm{B}(\hat{\mu}_3-\hat{\mu}_2),\quad \hat{\rho}_{\mathrm{ss}}=(\hat{\mu}_1+\hat{\mu}_2+\hat{\mu}_3)/3,   
\end{equation}
where $c_{\mathrm{A,B}}$ are real constants. Hence, $\hat{\mu}_{1,2,3}$ are stationary states of $\Lambda$ and linearly independent. This last property is better appreciated inverting (\ref{deco_modes}):
\begin{equation}
\hat{\mu}_1=\hat{\rho}_{\mathrm{ss}}+\frac{2}{3}(\hat{R}_\mathrm{A}/c_\mathrm{A}),\quad  \hat{\mu}_2=\hat{\rho}_{\mathrm{ss}}-\frac{1}{3}(\hat{R}_\mathrm{A}/c_\mathrm{A})-\frac{1}{2}(\hat{R}_\mathrm{B}/c_\mathrm{B}),\quad \hat{\mu}_3=\hat{\rho}_{\mathrm{ss}}-\frac{1}{3}(\hat{R}_\mathrm{A}/c_\mathrm{A})+\frac{1}{2}(\hat{R}_\mathrm{B}/c_\mathrm{B}).   
\end{equation}
which shows that they are an independent linear combination (non-zero determinant of the coefficients arranged in columns) of linearly independent vectors (the linear independence of  $\hat{\rho}_{\mathrm{ss}}$, $\hat{R}_\mathrm{A}/c_\mathrm{A}$ and $\hat{R}_\mathrm{B}/c_\mathrm{B}$ is guaranteed by the assumption that they are eigenmatrices of $\Lambda$). We can proceed in the same line for the left eigenmatrices, obtaining:
\begin{equation}\label{projectors}
\hat{P}_1=\frac{\mathbb{1}}{3}+c_\mathrm{A}\hat{L}_\mathrm{A},\quad  \hat{P}_2=\frac{\mathbb{1}}{3}-\frac{c_\mathrm{A}\hat{L}_\mathrm{A}}{2}   -c_\mathrm{B}\hat{L}_\mathrm{B},\quad \hat{P}_3=\frac{\mathbb{1}}{3}-\frac{c_\mathrm{A}\hat{L}_\mathrm{A}}{2}  +c_\mathrm{B}\hat{L}_\mathrm{B},
\end{equation}
which satisfy:
\begin{equation}\label{properties1}
\hat{P}_1+  \hat{P}_2  +\hat{P}_3=\mathbb{1}, \quad \text{Tr}[\hat{P}_j\hat{\mu}_k]=\delta_{jk} \,\,\forall j,k,
\end{equation}
as follows by definition. Moreover, they must satisfy 
\begin{equation}\label{properties2}
\hat{P}_j\geq 0\,\,\forall j    
\end{equation}
to guarantee that, whatever is the initial condition, $\mathcal{P}\hat{\rho}_{0}$ will be a positive semi-definite matrix. Indeed, from properties (\ref{properties1})-(\ref{properties2}), it follows that:
\begin{equation}\label{probabilistic_mixture}
\mathcal{P}\hat{\rho}_0=p_1^0\hat{\mu}_1 +p_2^0\hat{\mu}_2+p_3^0\hat{\mu}_3 ,  
\end{equation}
with $p_j^0=\text{Tr}[\hat{P}_j\hat{\rho}_0]$, and thus $p_j\geq 0$ and $p^0_1+p^0_2+p^0_3=1$. Therefore, $\mathcal{P}\hat{\rho}_0$ is a probability mixture of three stationary states, which motivates the term extreme metastable states (EMSs) \cite{Macieszczak2016}. We now address the question of whether $\hat{\mu}_{1,2,3}$ actually break the discrete time-translation symmetry of $\mathcal{E}$. Indeed, the result is that if:
\begin{equation}\label{condition_constants}
r=\frac{c_\mathrm{A}}{c_\mathrm{B}}=\frac{2}{\sqrt{3}}    
\end{equation}
then:
\begin{equation}\label{cyclic}
\mathcal{E}\hat{\mu}_1=\hat{\mu}_2, \quad \mathcal{E}\hat{\mu}_2=\hat{\mu}_3, \quad \mathcal{E}\hat{\mu}_3=\hat{\mu}_1,% \quad \mathcal{E}\hat{\mu}_1=\hat{\mu}_2.
\end{equation}
Notice that our initial assumption that $\Lambda \hat{R}_1=\hat{R}_1$ implies that $\nu_1=2\pi/3$. Then $\mathcal{E}R_1=(-1/2+i\sqrt{3}/2)\hat{R}_1$, and together with (\ref{condition_constants}) we obtain:
\begin{equation}
\mathcal{E}(\hat{R}_\mathrm{A}/c_\mathrm{A})= -\frac{1}{2}(\hat{R}_\mathrm{A}/c_\mathrm{A})- \frac{3}{4}(\hat{R}_\mathrm{B}/c_\mathrm{B}), \quad  \mathcal{E}(\hat{R}_\mathrm{B}/c_\mathrm{B})= (\hat{R}_\mathrm{A}/c_\mathrm{A})- \frac{1}{2}(\hat{R}_\mathrm{B}/c_\mathrm{B}),
\end{equation}
which can be used to obtain the cyclic relation (\ref{cyclic}).

\subsection{Approximate decomposition in metastable period-tripled lobes} 

\begin{figure}[t]
\includegraphics[width=1\linewidth]{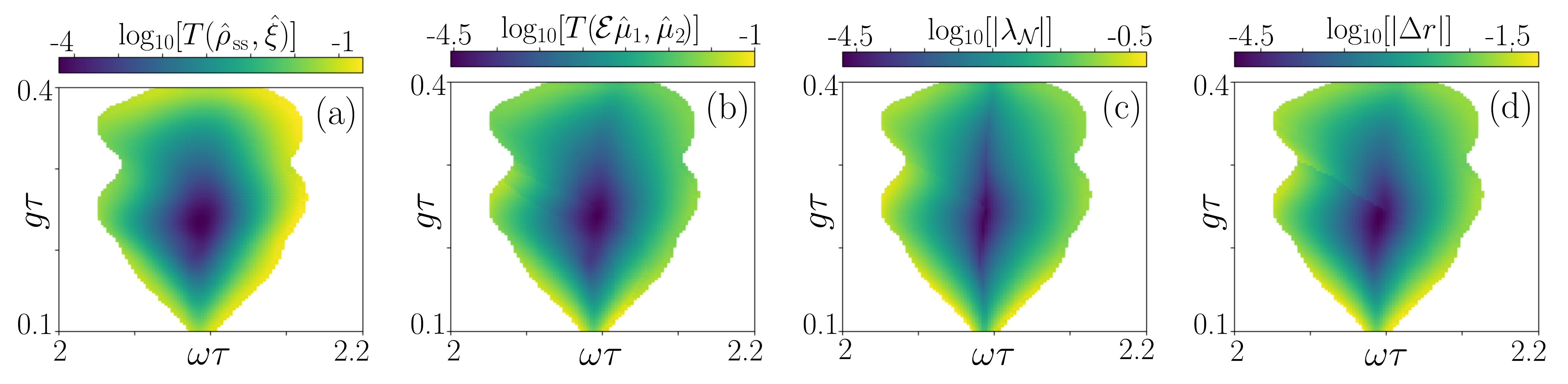}\\
\caption{Characterization of the main approximations in the metastable period-tripled regime for $N=30$. The plots are restricted to the region in which the dominant eigenmodes are the complex conjugate ones leading to period-tripling. All plots are in logarithmic scale. (a) Trace distance between the actual stationary state and its approximation given by Eq. (\ref{approximation_lobes}). (b) Trace distance between $\mathcal{E}\hat{\mu}_1$ and $\hat{\mu}_2$. (c) Absolute value of the largest negative eigenvalue of $\hat{P}_1$, denoted by $\lambda_\mathcal{N}$. (d) Absolute value of the difference between the actual value of $r$ and the theoretical one, i.e. $2/\sqrt{3}$. Here $\Delta r=r-2/\sqrt{3}$.}\label{fig_checks}
\end{figure}

In our system $\delta_1\tau\ll1$, $\Gamma_1\tau\ll1$, although they are non-zero. This means that the results of the previous subsection do not apply exactly but only in an approximate way. The most straightforward way to obtain an approximation for $\hat{\mu}_{1,2,3}$ is to use the spectral decomposition of $\hat{R}_{\mathrm{A,B}}$. Indeed, we can choose:
\begin{equation}
\hat{\mu}_2=\frac{1}{c_\mathrm{B}}\sum_{b_j<0} b_j|B_j\rangle\langle B_j|, \quad   \hat{\mu}_3=\frac{1}{c_\mathrm{B}}\sum_{b_j>0} b_j|B_j\rangle\langle B_j|, \quad c_\mathrm{B}=\sum_{b_j>0}b_j,
\end{equation}
\begin{equation}
\hat{\mu}_1=\frac{1}{c_\mathrm{A}}\sum_{a_j>0} a_j|A_j\rangle\langle A_j|,  \quad c_\mathrm{A}=\sum_{a_j>0}a_j.
\end{equation}
which ensures that $\hat{\mu}_{1,2,3}$ are {\it bona fide} density matrices. However, then we have that some of the following relations hold under the approximation sign:
\begin{equation}\label{approximation_lobes}
\hat{R}_\mathrm{A}\approx c_\mathrm{A}(\hat{\mu}_1-\hat{\mu}_2/2-\hat{\mu}_3/2),\quad  \hat{R}_\mathrm{B}=c_\mathrm{B}(\hat{\mu}_3-\hat{\mu}_2),\quad \hat{\rho}_{\mathrm{ss}}\approx(\hat{\mu}_1+\hat{\mu}_2+\hat{\mu}_3)/3=\xi,   
\end{equation}
and their accuracy  needs to be checked. Moreover, we will also find that Eqs. (\ref{probabilistic_mixture}) and (\ref{condition_constants}) also apply only approximately, while $\text{Tr}[\hat{P}_j\hat{\mu}_k]\approx\delta_{jk}$ and they are only approximately positive. Therefore, the accuracy of such approximations need to be characterized for the metastable region. This is done in Fig. \ref{fig_checks}, in which we characterize the main approximations: (a) the trace distance between the actual stationary state and its approximation; (b) the trace distance betweeen $\mathcal{E}\hat{\mu}_1$ and $\hat{\mu}_2$; (c) the largest negative eigenvalue of $\hat{P}_1$ denoted by $\lambda_\mathcal{N}$; and (d) the deviation of $r$ from $2/\sqrt{3}$, denoted by $\Delta r$. Comparing these results with those of Fig. \ref{fig_meta} (a) of the main text, we observe that the larger is $\Gamma_*/\Gamma_1$, the better are these approximations. We conclude that these approximate results are in good agreement with the exact ones for the relevant metastable region.

\subsection{Classical relaxation dynamics}

A remarkable result is that not only $\mathcal{P}\hat{\rho}_0$ can be approximated as a probabilistic mixture of three metstable states, but the entire relaxation dynamics of $\mathcal{P}\hat{\rho}_{3n}$ can be described by a classical three state stochastic process. In order to show this, we will make use of the rescaled Hermitian modes and their left partners which we define as:
\begin{equation}
\hat{R}'_\mathrm{A,B}=\hat{R}_\mathrm{A,B}/c_{\mathrm{A,B}}, \quad   \hat{L}'_\mathrm{A,B}=c_{\mathrm{A,B}}\hat{L}_\mathrm{A,B}.  
\end{equation}
Rewriting the stroboscopic dynamics given by Eq. (\ref{stroboscopic_dynamics}) in terms of the rescaled Hermitian modes we obtain:
\begin{equation}
\mathcal{P}\hat{\rho}_{3n}=\hat{\rho}_{\mathrm{ss}}+[A'\cos(\delta_1 n\tau)+r\,B'\sin (\delta_1  n\tau) ]\hat{R}'_\mathrm{A} e^{-\Gamma_1  n\tau} +[B'\cos(\delta_1 n\tau)-A'\sin (\delta_1  n\tau)/r ]\hat{R}'_\mathrm{B} e^{-\Gamma_1  n\tau} ,
\end{equation}
where $A'=\text{Tr}[\hat{L}_\mathrm{A}'\hat{\rho}_0]$ and $B'=\text{Tr}[\hat{L}_\mathrm{B}'\hat{\rho}_0]$. Then, we make use of the expression of the eigenmodes in terms of the EMSs given in Eq. (\ref{approximation_lobes}) to obtain:
\begin{equation}\label{final_relaxation}
\mathcal{P}\hat{\rho}(3n\tau)\approx p_1(n\tau)\hat{\mu}_1+  p_2(n\tau)\hat{\mu}_2+  p_3(n\tau)\hat{\mu}_3
\end{equation}
where the $p_j(n\tau)$ are given   in terms of only the initial conditions $p_j^0$ and the rates $\Gamma_1$ and $\delta_1$:
\begin{equation}
\begin{split}
p_1(n\tau)&=\frac{p_1^0}{3}(1+2\cos (\delta_1 n\tau)e^{-\Gamma_1n\tau})+\frac{p_3^0}{3}(1-[\cos (\delta_1 n\tau)-\sqrt{3}\sin (\delta_1 n\tau)]e^{-\Gamma_1n\tau})\\
&+\frac{p_2^0}{3}(1-[\cos (\delta_1 n\tau)+\sqrt{3}\sin (\delta_1 n\tau)]e^{-\Gamma_1n\tau}),\label{solution1}\\
\end{split}
\end{equation}
\begin{equation}
\begin{split}
p_2(n\tau)&=\frac{p_2^0}{3}(1+2\cos (\delta_1 n\tau)e^{-\Gamma_1n\tau})+\frac{p_1^0}{3}(1-[\cos (\delta_1 n\tau)-\sqrt{3}\sin (\delta_1 n\tau)]e^{-\Gamma_1n\tau})\\
&+\frac{p_3^0}{3}(1-[\cos (\delta_1 n\tau)+\sqrt{3}\sin (\delta_1 n\tau)]e^{-\Gamma_1n\tau}),\label{solution2}
\end{split}
\end{equation}
\begin{equation}
\begin{split}
p_3(n\tau)&=\frac{p_3^0}{3}(1+2\cos (\delta_1 n\tau)e^{-\Gamma_1n\tau})+\frac{p_2^0}{3}(1-[\cos (\delta_1 n\tau)-\sqrt{3}\sin (\delta_1 n\tau)]e^{-\Gamma_1n\tau})\\
&+\frac{p_1^0}{3}(1-[\cos (\delta_1 n\tau)+\sqrt{3}\sin (\delta_1 n\tau)]e^{-\Gamma_1n\tau}).\label{solution3}\\
\end{split}
\end{equation}
Notice that here we have used the approximations $c_\mathrm{A}/c_\mathrm{B}\approx 2/\sqrt{3}$, Eq. (\ref{approximation_lobes}), and the expression of $p^0_j$ in terms of $A'$ and $B'$, which can be obtained from their definition $p_j^0=\text{Tr}[\hat{P}_j\hat{\rho}_0]$ and read:
\begin{equation}
p_1^0=\frac{1}{3}+A', \quad  p_2^0=\frac{1}{3}-\frac{A'}{2}-B',\quad     p_3^0=\frac{1}{3}-\frac{A'}{2}+B'.
\end{equation}
The use of these approximations is the reason why we write Eq. (\ref{final_relaxation}) under the approximate sign. Finally, we recognize (i.e. we can check) that Eqs. (\ref{solution1})-(\ref{solution3}) are indeed the solution at discrete time steps $t=n\tau$ of the classical stochastic process given by: 
\begin{align}
\frac{d}{dt}p_1&=-\frac{2\Gamma_1}{3}p_1+\bigg(\frac{\Gamma_1}{3}-\frac{\delta_1}{\sqrt{3}} \bigg)p_2+ \bigg(\frac{\Gamma_1}{3}+\frac{\delta_1}{\sqrt{3}} \bigg)p_3,\label{classical_process1}\\
\frac{d}{dt}p_2&=-\frac{2\Gamma_1}{3}p_2+\bigg(\frac{\Gamma_1}{3}-\frac{\delta_1}{\sqrt{3}} \bigg)p_3+ \bigg(\frac{\Gamma_1}{3}+\frac{\delta_1}{\sqrt{3}} \bigg)p_1,\label{classical_process2}\\
\frac{d}{dt}p_3&=-\frac{2\Gamma_1}{3}p_3+\bigg(\frac{\Gamma_1}{3}-\frac{\delta_1}{\sqrt{3}} \bigg)p_1+ \bigg(\frac{\Gamma_1}{3}+\frac{\delta_1}{\sqrt{3}} \bigg)p_2,\label{classical_process3}
\end{align}
with initial contitions $p_j^0$. Notice that the condition for these equations to represent a classical stochastic process is that:
\begin{equation}\label{condition_classical_process}
\frac{\Gamma_1}{|\delta_1|}\geq \sqrt{3},    
\end{equation}
since the off-diagonal rates need to be positive. From Fig. \ref{fig_meta} (b) of the main text, we see that this condition is widely satisfied. Then, Eqs. (\ref{classical_process1})-(\ref{classical_process3}) guarantee that $p_j(t)$ can be regarded as probabilities at all times. This process generalizes to three states the effective classical dynamics disclosed in Refs. \cite{Macieszczak2016,Rose2016}. We observe that when considering three states, there is the possibility to have a very slow oscillation in the long-time relaxation as described by the terms proportional to $\delta_1$. 

\subsection{Stationary current}

The classical stochastic process given by Eqs. (\ref{classical_process1})-(\ref{classical_process3}) can be written in matrix form:
\begin{equation}
\frac{d}{dt}\vec{p}=W\vec{p}    
\end{equation}
where $\vec{p}=(p_1,p_2,p_3)^T$ and
\begin{equation}
W=
 \begin{pmatrix}
  -\frac{2\Gamma_1}{3} & & \frac{\Gamma_1}{3}-\frac{\delta_1}{\sqrt{3}} & & \frac{\Gamma_1}{3}+\frac{\delta_1}{\sqrt{3}} \\
  & & & & \\
  \frac{\Gamma_1}{3}+\frac{\delta_1}{\sqrt{3}} &  & -\frac{2\Gamma_1}{3} & & \frac{\Gamma_1}{3}-\frac{\delta_1}{\sqrt{3}} \\
    & & & & \\
  \frac{\Gamma_1}{3}-\frac{\delta_1}{\sqrt{3}} & & \frac{\Gamma_1}{3}+\frac{\delta_1}{\sqrt{3}} & & -\frac{2\Gamma_1}{3}
 \end{pmatrix}.    
\end{equation}
The matrix $W$ satisfies the following properties that guarantee it to be a classical stochastic process: (i) the sum of each of the columns is zero, i.e. $\sum_{i=1}^3 W_{ij}=0$ $\forall j$, which guarantees conservation of probability; (ii) the off-diagonal elements are positive semi-definite (if $\Gamma_1\geq\sqrt{3}|\delta_1|$), i.e. $W_{ij}\geq0$ if $i\neq j$, which guarantees the $p_j$'s to remain positive. Moreover, a third property (iii) is that the sum of the elements of the same row is zero, i.e. $\sum_{j=1}^3 W_{ij}=0$ $\forall i$, which indicates that the stationary state is uniform, i.e. $p_1^\infty=p_2^\infty=p_3^\infty=1/3$. Despite the stationary state is uniform, it displays non-zero currents and thus it is a {\it non-equilibrium} stationary state. The stationary current from state $i$ to state $j$ is given by:
\begin{equation}
J_{ij} =p_i^\infty W_{ji}-   p_{j}^\infty W_{ij}. 
\end{equation}
From which we find that
\begin{equation}
J=J_{12}=J_{23}=J_{31}=\frac{2\delta_1}{3\sqrt{3}}.    
\end{equation}
Thus if $\delta_1>0$ there is a stationary clockwise probability current, while if $\delta_1<0$ there is a stationary anti-clockwise probability current.

\section{Period-5 metastable oscillations}

\begin{figure}[t]
\includegraphics[width=0.8\linewidth]{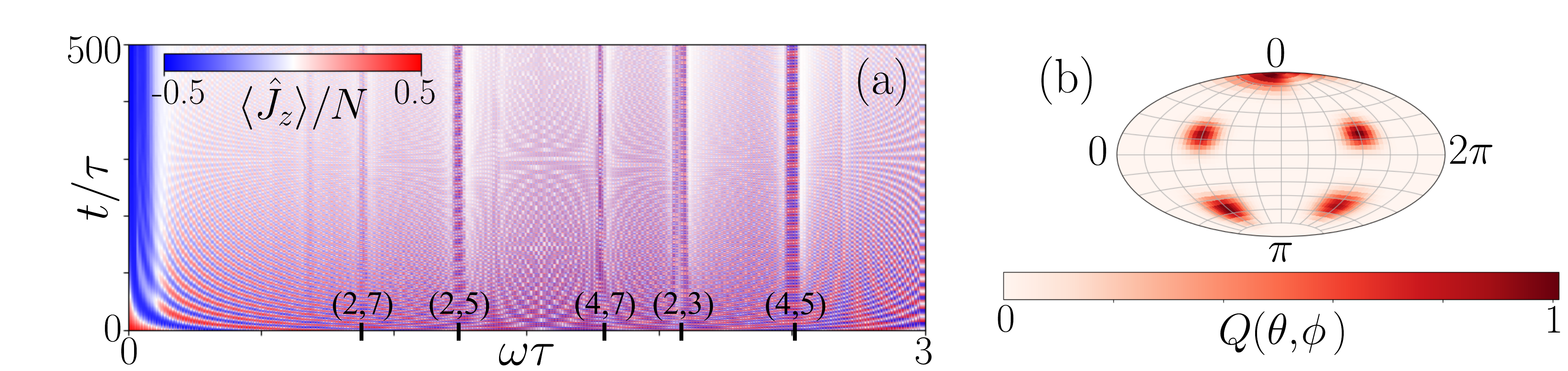}\\
\caption{(a) Magnetization dynamics varying the Rabi frequency and for an exemplary cut at $g\tau=0.18$, with the initial condition $|J,J\rangle$ and system size $N=70$.  In the horizontal axis we have marked the most prominent metastable resonances, where the notation (p,q) resonance stands for a resonance around the frequency $\omega=p\pi/q$. (b) Aitoff projection of the Husimi Q function for the stationary state in the (4,5) resonance. The parameters are $\omega\tau=4\pi/5$, $g\tau=0.18$ and $N=70$.}\label{fig_baldosa5}
\end{figure}

\subsection{Long-time dynamics}

Besides the prominent $p=2$, $q=3$ region in which we have focused, we can also find other types of metastable resonances. In particular, a second important one is that occurring for $p=4$ and $q=5$. While signatures of this resonance in the stationary state are already evident for $N=30$, i.e. $\text{Tr}[\hat{\rho}^2_{\mathrm{ss}}]\sim1/5$ and $\hat{\rho}_{\mathrm{ss}}$ made of 5 disjoint lobes, this becomes more important in systems with larger sizes. For instance, in Fig.  \ref{fig_baldosa5} (a), we exemplify the dynamics for $N=70$ varying $\omega\tau$ and for a cut at $g \tau=0.18$, similarly to what we have done in the main text. We can appreciate both the transition from overdamped to underdamped dynamics as well as the presence of metastable resonances for several combinations of $(p,q)$, the (4,5) resonance being particularly clear. In panel (b) we plot the stationary state for a point inside the $(4,5)$ resonance, finding that it displays five almost disjoint lobes, as anticipated in the main text. 

In the $(4,5)$ resonance we find the long-time dynamics to be accurately described by:
\begin{equation}\label{longtime_approx5}
\begin{split}
\hat{\rho}_n\approx\mathcal{P}_5\hat{\rho}_n= &\hat{\rho}_{\mathrm{ss}}+\text{Tr}[\hat{L}_1\hat{\rho}_0]\hat{R}_1 e^{i\nu_1 n\tau -\gamma_1n\tau}+  \text{Tr}[\hat{L}^\dagger_1\hat{\rho}_0]\hat{R}^\dagger_1 e^{-i\nu_1 n\tau -\gamma_1n\tau}\\
&+\text{Tr}[\hat{L}_2\hat{\rho}_0]\hat{R}_2 e^{i\nu_2 n\tau -\gamma_2n\tau}+  \text{Tr}[\hat{L}^\dagger_2\hat{\rho}_0]\hat{R}^\dagger_2 e^{-i\nu_2 n\tau -\gamma_2n\tau}
\end{split}
\end{equation}
Notice that we stick to the following criteria for the definition of the eigenvalues and eigenmatrices: $\nu_1>0$, $\nu_2>0$, $\text{Tr}[\hat{L}_1 \hat{R}_1]=1$ and $\text{Tr}[\hat{L}_2 \hat{R}_2]=1$. Moreover, in this resonance, we typically find that $\gamma_1\sim\gamma_2$, while $\nu_1\tau\sim 4\pi/5$ and $\nu_2\tau\sim 2\pi/5$. As in the period-tripled case, we will work with the Hermitian counterparts of the long-lived eigenmodes:
\begin{equation}\label{Hermitian_partners5}
\begin{split}
&\hat{R}_\mathrm{A}=\frac{\hat{R}_1+\hat{R}_1^\dagger}{2}, \quad \hat{R}_\mathrm{B}=\frac{\hat{R}_1-\hat{R}_1^\dagger}{2i},\quad
\hat{R}_\mathrm{C}=\frac{\hat{R}_2+\hat{R}_2^\dagger}{2}, \quad \hat{R}_\mathrm{D}=\frac{\hat{R}_2-\hat{R}_2^\dagger}{2i},\\
&\hat{L}_\mathrm{A}=\hat{L}_1+\hat{L}_1^\dagger,\quad 
\hat{L}_\mathrm{B}=i(\hat{L}_1-\hat{L}_1^\dagger),\quad
\hat{L}_\mathrm{C}=\hat{L}_2+\hat{L}_2^\dagger,\quad 
\hat{L}_\mathrm{D}=i(\hat{L}_2-\hat{L}_2^\dagger),
\end{split}
\end{equation}
which  satisfy $\text{Tr}[\hat{L}_j \hat{R}_k]=\delta_{jk}$  with $j,k\in\{\mathrm{A,B,C,D}\}$. Since they are Hermitian and traceless, we can decompose them as the substraction of two physical states:
\begin{equation}
\hat{R}_\mathrm{X}=\sum_{j=1}^{2J+1} x_j|X_j\rangle\langle X_j|,\quad \sum_{j=1}^{2J+1} x_j=0,\quad \text{with}\quad x=a,b,c,d.
\end{equation}
Importantly, this decomposition allows us to define the following constants:
\begin{equation}
c_\mathrm{X}=\sum_{x_j>0}x_j,\quad  \text{with}\quad x=a,b,c,d. 
\end{equation}
\subsection{Formal decomposition in period-5 lobes}
We consider now the period-5 map $\Lambda_5=\mathcal{E}^5$, which displays the same eigenmatrices but whose eigenvalues are a factor five those of $\mathcal{E}$. We suppose that this map displays a gap closure in the region $p=4$, $q=5$, in which $\Gamma_{1,2}=0$ and $\delta_{1,2}=0$, and we propose a decomposition of the involved modes in terms of a set of period-5 states. Numerical observation leads us to propose the following decomposition:
\begin{equation}\label{P5_ss}
\hat{\rho}_{\mathrm{ss}}=\frac{1}{5}(\hat{\mu}_1+\hat{\mu}_2+\hat{\mu}_3+\hat{\mu}_4+\hat{\mu}_5),    
\end{equation}
\begin{equation}\label{P5_mode1}
\hat{R}_{\mathrm{A}}=\frac{c_\mathrm{A}}{5}(3\hat{\mu}_1+\hat{\mu}_2+\hat{\mu}_5)-\frac{c_{\mathrm{A}}}{2}(\hat{\mu}_3+\hat{\mu}_4),    
\end{equation}
\begin{equation}\label{P5_mode2}
\hat{R}_{\mathrm{B}}=\frac{c_\mathrm{B}}{5}(3\hat{\mu}_5+2\hat{\mu}_4-3\hat{\mu}_2-2\hat{\mu}_3),    
\end{equation}
\begin{equation}\label{P5_mode3}
\hat{R}_{\mathrm{C}}=\frac{c_\mathrm{C}}{5}(3\hat{\mu}_1+\hat{\mu}_3+\hat{\mu}_4)-\frac{c_\mathrm{C}}{2}(\hat{\mu}_2+\hat{\mu}_5),    
\end{equation}
\begin{equation}\label{P5_mode4}
\hat{R}_{\mathrm{D}}=\frac{c_\mathrm{D}}{5}(2\hat{\mu}_2+3\hat{\mu}_4-3\hat{\mu}_3-2\hat{\mu}_5).    
\end{equation}
We then obtain the following expressions for the different lobes:
\begin{equation}
\hat{\mu}_1=\hat{\rho}_{\mathrm{ss}}+\frac{2}{3}(\hat{R}_\mathrm{A}'+\hat{R}_\mathrm{C}'),    
\end{equation}
\begin{equation}
\hat{\mu}_2=\hat{\rho}_{\mathrm{ss}}+\frac{1}{546}(104\hat{R}_\mathrm{A}'-315\hat{R}_\mathrm{B}'-286\hat{R}_\mathrm{C}'+210\hat{R}_\mathrm{D}'),
\end{equation}
\begin{equation}
\hat{\mu}_3=\hat{\rho}_{\mathrm{ss}}+\frac{1}{546}(-286\hat{R}_\mathrm{A}'-210\hat{R}_\mathrm{B}'+104\hat{R}_\mathrm{C}'-315\hat{R}_\mathrm{D}'),
\end{equation}
\begin{equation}
\hat{\mu}_4=\hat{\rho}_{\mathrm{ss}}+\frac{1}{546}(-286\hat{R}_\mathrm{A}'+210\hat{R}_\mathrm{B}'+104\hat{R}_\mathrm{C}'+315\hat{R}_\mathrm{D}'),
\end{equation}
\begin{equation}
\hat{\mu}_5=\hat{\rho}_{\mathrm{ss}}+\frac{1}{546}(104\hat{R}_\mathrm{A}'+315\hat{R}_\mathrm{B}'-286\hat{R}_\mathrm{C}'-210\hat{R}_\mathrm{D}'),
\end{equation}
where we have defined $\hat{R}_\mathrm{X}'=\hat{R}_\mathrm{X}/c_\mathrm{X}$, with $\mathrm{X}=\mathrm{A,B,C,D}$. Similarly as in the period-tripled case, we also define their left partners as:
\begin{equation}
\hat{P}_1=\frac{1}{5}(\mathbb{1}+3\hat{L}'_\mathrm{A}+3\hat{L}'_\mathrm{C}),    
\end{equation}
\begin{equation}
\hat{P}_2=\frac{1}{5}(\mathbb{1}+\hat{L}'_\mathrm{A}-3\hat{L}'_\mathrm{B}+2\hat{L}'_\mathrm{D})-\frac{\hat{L}'_\mathrm{C}}{2},    
\end{equation}
\begin{equation}
\hat{P}_3=\frac{1}{5}(\mathbb{1}-2\hat{L}'_\mathrm{B}+\hat{L}'_\mathrm{C}-3\hat{L}'_\mathrm{D})-\frac{\hat{L}'_\mathrm{A}}{2},    
\end{equation}
\begin{equation}
\hat{P}_4=\frac{1}{5}(\mathbb{1}+2\hat{L}'_\mathrm{B}+\hat{L}'_\mathrm{C}+3\hat{L}'_\mathrm{D})-\frac{\hat{L}'_\mathrm{A}}{2},    
\end{equation}
\begin{equation}
\hat{P}_5=\frac{1}{5}(\mathbb{1}+\hat{L}'_\mathrm{A}+3\hat{L}'_\mathrm{B}-2\hat{L}'_\mathrm{D})-\frac{\hat{L}'_\mathrm{C}}{2},    
\end{equation}
where we have defined $\hat{L}_\mathrm{X}'=c_\mathrm{X} \hat{L}_\mathrm{X}$, with $\mathrm{X}=\mathrm{A,B,C,D}$. These operators satisfy:
\begin{equation}
\sum_{j=1}^5 \hat{P}_j=\mathbb{1},\quad \text{Tr}[\hat{P}_i\hat{\mu}_j]=\delta_{ij}.    
\end{equation}
Here we notice that the positivity of $\hat{P}_i$ and $\hat{\mu}_i$ is interrelated. If there was an actual gap closure and thus $\hat{\mu}_i$ were true stationary states we would necessarily have $\hat{P}_i\geq 0$.

\begin{figure}[t]
\includegraphics[width=1\linewidth]{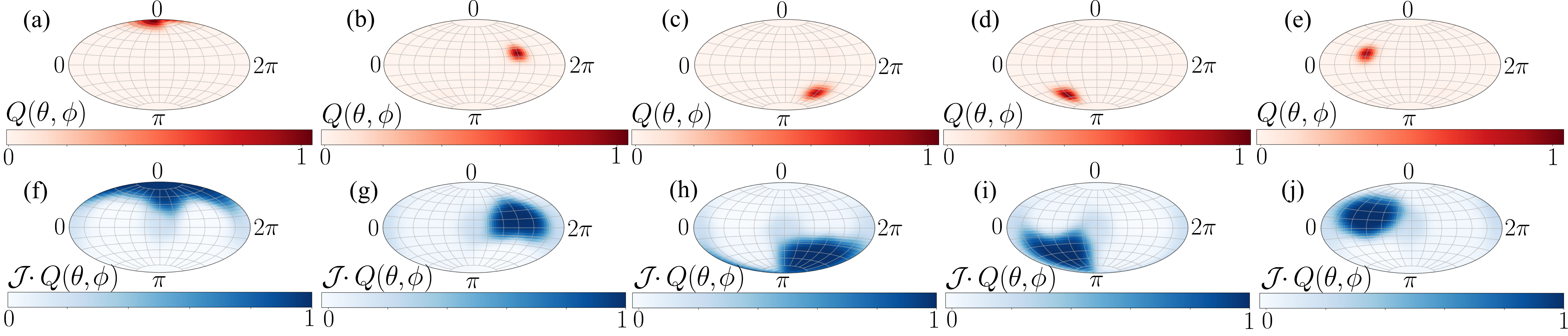}\\
\caption{Aitoff projection of the Husimi Q function of the  period-5 lobes and their left partners for  N=70, $g\tau=0.18$, $\omega\tau=4\pi/5$.  (a)-(e) Metastable lobes: $\hat{\mu}_1/5$ to $\hat{\mu}_5/5$, respectively. (f)-(j) Left projectors: $\hat{P}_1$ to $\hat{P}_5$, respectively. For these parameters the leading decay rates and frequencies are given by: $\gamma_1\tau=1.3\cdot 10^{-5}$, $\nu_1\tau=0.25133$; $\gamma_2\tau=3.2\cdot 10^{-5}$, $\nu_2\tau=1.2566$;  while the next mode has $\gamma_*\tau=3.9\cdot 10^{-4}$, $\nu_*\tau=0$, with $*=5$ as there are two pairs of long-lived metastable modes. In this case we have a ratio of $\gamma_*/\gamma_2=12.2$ between the largest rate of the metastable manifold and the next one. Moreover, we obtain the following values for the figures of merit:  $T(\mathcal{E}\hat{\mu}_j,\hat{\mu}_{j+2})\approx 0.033$, while  the smallest eigenvalue of $\hat{P}_j$ takes values around $\lambda_\mathcal{N}\approx -0.02$. Finally due to the way in which we have defined the metastable lobes and projectors, we have that the definitions in Eqs. (\ref{P5_ss}) to (\ref{P5_mode4}) and $\text{Tr}[\hat{P}_j\hat{\mu}_k]=\delta_{jk}$ are satisfied exactly (and thus the corresponding trace distances are zero). }\label{fig_lobes5}
\end{figure}

\subsection{Approximate decomposition in period-5 lobes}

For the period-5 case we follow a slightly different strategy to define the approximate EMSs than in the period-tripled case. Instead of defining them from the spectral decomposition of the Hermitian partners of the eigenmodes, we define them through the relations given in Eqs. (\ref{P5_ss}) to (\ref{P5_mode4}). In principle, if there was an actual gap closure both ways would provide equivalent results. However, in practice this means that instead of $\hat{\mu}_j$ being {\it bona fide} states and equations (\ref{approximation_lobes}) holding approximately, we now have it in the other way around: equations  (\ref{P5_ss}) to (\ref{P5_mode4}) hold exactly, however $\hat{\mu}_j$ and their left partners display small corrections to positivity. The reason why here we proceed in this different way is that, due to the increased complexity of the MM, this is the most straightforward manner of isolating the metastable lobes. Nevertheless, we recall that if the approximation is good such differences remain small. As we will show now, these metastable approximations also work well for the period-5 case.

In Fig. \ref{fig_lobes5} we exemplify this lobe decomposition, from which we can appreciate that the EMSs correspond to the lobes making up the stationary state shown in Fig. \ref{fig_baldosa5} (b). Here, we also check for the period-5 cyclic relation connecting the EMSs, which we numerically find to be:
\begin{equation}\label{period5_cyclic}
\mathcal{E}\hat{\mu}_1\approx\hat{\mu}_{3}, \quad  \mathcal{E}\hat{\mu}_2\approx\hat{\mu}_{4}, \quad \mathcal{E}\hat{\mu}_3\approx\hat{\mu}_{5}, \quad\mathcal{E}\hat{\mu}_4\approx\hat{\mu}_{1} \quad \mathcal{E}\hat{\mu}_5\approx\hat{\mu}_{2}.
\end{equation}
Regarding the corrections to these approximations, we find the smallest eigenvalue of the $\hat{P}_j$ to take values around $\lambda_\mathcal{N}\approx -0.02$, while trace distances for the cyclic relation take values around $T(\mathcal{E}\hat{\mu}_j,\hat{\mu}_{j+2})\approx 0.033$. Comparing this with the results for the period-tripled case shown in Fig. \ref{fig_checks}, we see that the corrections in the period-5 case are generally larger. This is in agreement with the fact that the spectral gap between the eigenvalues of the MM and the smallest ones outside it is smaller compared to the period-tripled case, taking the value  $\gamma_*/\gamma_2=12.2$ for the chosen parameters. As we shall see, although the corrections for these approximations are larger, we still find good agreement when looking at the dynamics (see  next subsection). This good agreement can be in part attributed to the immediate leading eigenmodes outside the MM  not playing an important role for the z- y-components of the magnetization dynamics.

\subsection{Approximate metastable dynamics}

\begin{figure}[t]
\includegraphics[width=1\linewidth]{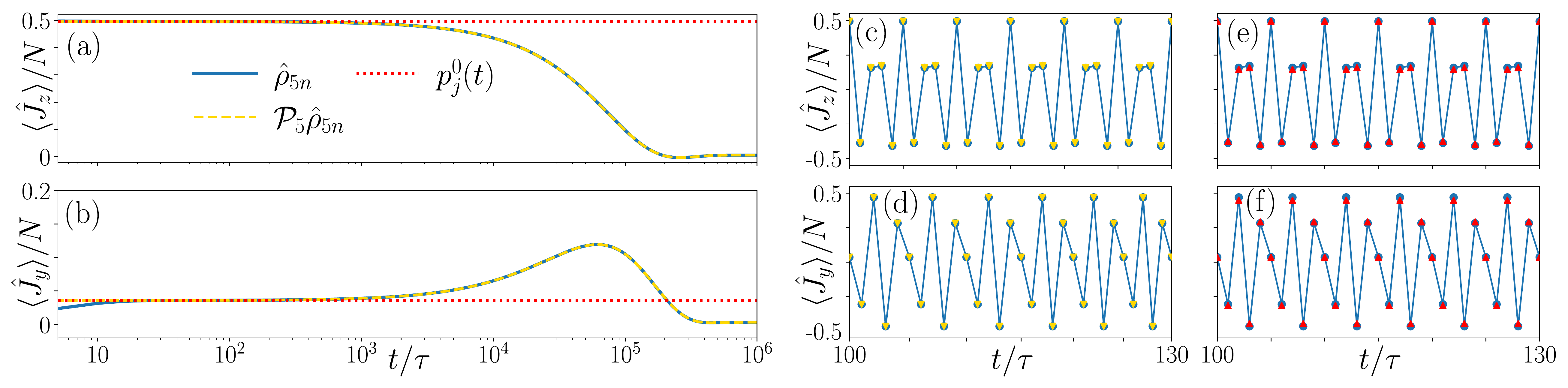}\\
\caption{(a)-(b) Stroboscopic period-5 dynamics for the magnetization in the $z$ and $y$ direction. Exact results in blue solid lines. In golden dashed lines, results according to the projection onto the MM as given by Eq. (\ref{longtime_approx5}). In red dotted lines, initial incoherent mixture as given by Eq. (\ref{period5_mixture}). Initial condition $|J,J\rangle$. Parameters: $N=70$, $g\tau=0.18$, $\omega\tau=4\pi/5$. (c)-(f) Zoom in of the oscillatory dynamics for an interval of time in the metastable plateau. (c)-(d) Comparison of the approximation given by Eq. (\ref{longtime_approx5}) (golden down triangles) with the exact results (blue circles). (e)-(f) Comparison of the approximation given by Eq. (\ref{period5_MM}) (red up triangles) with the exact results (blue circles). Same parameters and initial condition as in (a) and (b).}\label{fig_trajs}
\end{figure}

In this subsection we want to compare the exact dynamics with the approximate ones for the (4,5) resonance. In particular, a first level of approximation is given by Eq. (\ref{longtime_approx5}), in which we have neglected the contributions of modes outside the MM for long times. In Fig. \ref{fig_trajs} (a)-(d) we compare this approximation (in golden dashed lines and triangles) with the exact dynamics (blue solid lines and circles), finding that after a short initial transient both display excellent agreement. A further level of approximation consists in approximating the state of the system in the metastable plateau by the initial probabilistic mixture of EMSs:
\begin{equation}\label{period5_mixture}
\mathcal{P}_5\hat{\rho}_0\approx p_1^0\hat{\mu}_1+p_2^0\hat{\mu}_2+ p_3^0\hat{\mu}_3+p_4^0\hat{\mu}_4+p_5^0\hat{\mu}_5.
\end{equation}
where $p_j^0=\text{Tr}[\hat{P}_j\hat{\rho}_0]$. Indeed, the $z$ and $y$ components of the magnetization according to Eq. (\ref{period5_mixture}) are shown in Fig. \ref{fig_trajs} (a), (b) in red-dotted lines, finding excellent agreement within the metastable plateau, that is after an initial short transient and before the final decay takes place. Moreover, making use of Eq. (\ref{period5_cyclic}) we can approximate the period-5 dynamics inside the plateau as:
\begin{equation}\label{period5_MM}
\hat{\rho}_{5n+j}\approx p_1^0\hat{\mu}_{1+2j}+p_2^0\hat{\mu}_{2+2j}+p_3^0\hat{\mu}_{3+2j}+p_4^0\hat{\mu}_{4+2j}+p_5^0\hat{\mu}_{5+2j},    
\end{equation}
where $j=0,1,2,3,4$ and the index of the metastable states follows periodic bounary conditions, i.e. $\hat{\mu}_{5+k}=\hat{\mu}_{\text{mod}(5+k,5)}$ with $k\geq 1$. In Fig. \ref{fig_trajs} (e) and (f) we compare the exact oscillatory dynamics (blue circles) with the approximation given in Eq. (\ref{period5_MM}) (red triangles). We notice that the agreement is not so good as for the same approximation done in the period-tripled case (main text), although differences between exact dynamics and this approximation are still reasonably small. The fact that this kind of approximation does not work so well as in the period-tripled case can be traced back to Eq. (\ref{period5_cyclic}) also not working so well (see also caption of Fig. \ref{fig_lobes5}), which  as we have already commented it can in turn be traced back to the spectral gap between the MM and the rest of eigenmodes being not so accentuated as in the $(2,3)$ resonance studied in the main text. 

In conclusion, besides quantitative differences in the level of precision of these approximations, we find our main results to  apply also for this case. The only result that we have not generalized to this higher-order resonance is a (possible) classical stochastic process describing the final relaxation in a stroboscopic picture.

\section{Increasing system size}

Finally, we want to illustrate what is the general effect of increasing system size on the metastable resonances. As stated in the main text, these metastable resonances are indeed also resonances in system size:  their emergence depends on system size and actually they do not display a smooth behavior when the thermodynamic limit is approached. We show this behavior in two different ways: by studying the lifetime of the dominant oscillatory mode with increasing system size and by showing how the shape and number of purity islands (associated to the metastable resonances) change with system size.

\subsection{Behavior of the lifetime of the dominant oscillation mode}

\begin{figure}[t]
\includegraphics[width=1\linewidth]{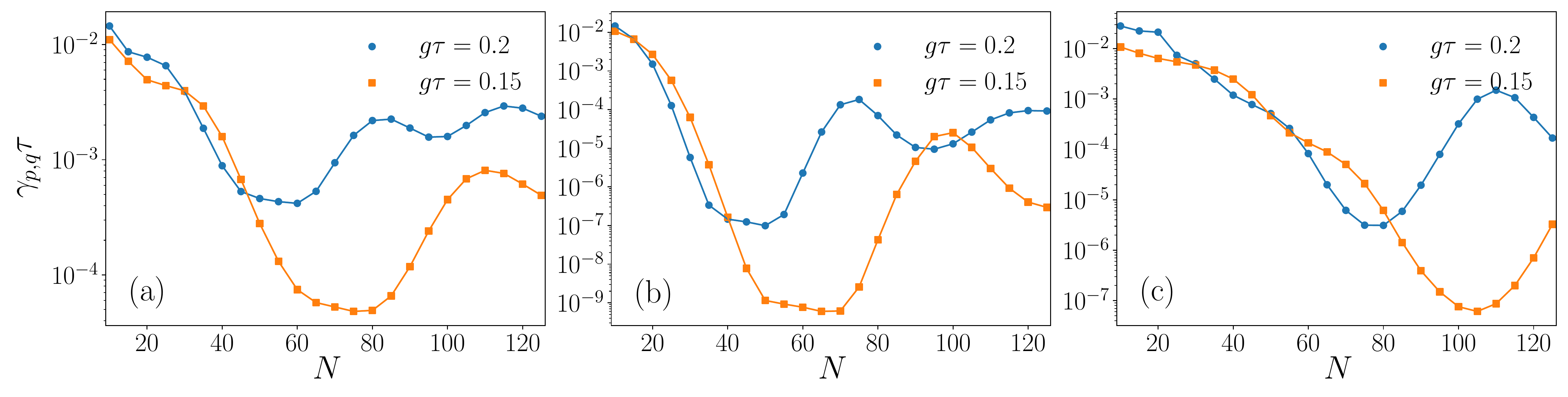}\\
\caption{Decay rate of the dominant oscillatory mode in the $(p,q)$ metastable resonance, $\gamma_{p,q}$, varying the system size and for $g\tau=0.2$ (blue circles) or $g\tau=0.15$ (orange squares). (a) $\omega=2\pi/5$. (b) $\omega=2\pi/3$.  (c) $\omega=4\pi/5$.}\label{fig_gapsizes}
\end{figure}

Here we explore the behavior of the lifetime of the dominant oscillatory mode (i.e. that with smallest decay rate) varying the system size and in different metastable resonances. In particular, in Fig. \ref{fig_gapsizes} we consider two values of $g\tau$ and we plot $\gamma_{p,q}$ for a Rabi frequency $\omega$ in: (a) the (2,5) resonance; (b) the (2,3) resonance; (c) the (4,5) resonance. In all the cases the results are qualitatively the same: the decay rate of the dominant oscillatory mode, i.e. that responsible of the main oscillation features, displays a non-monotonic behavior with $N$, and indeed it shows oscillations with system size. Thus, in contrast to dissipative phase transitions \cite{Minganti2018}, there is no evidence of spectral gap closure in the thermodynamic limit. Indeed, these results point out in the direction of considering these metastable resonances as resulting from a many-body resonance in which a large enough system size is needed, however further increasing system size can hinder the phenomenon.

\subsection{Behavior of the purity islands}

\begin{figure}[b]
\includegraphics[width=1\linewidth]{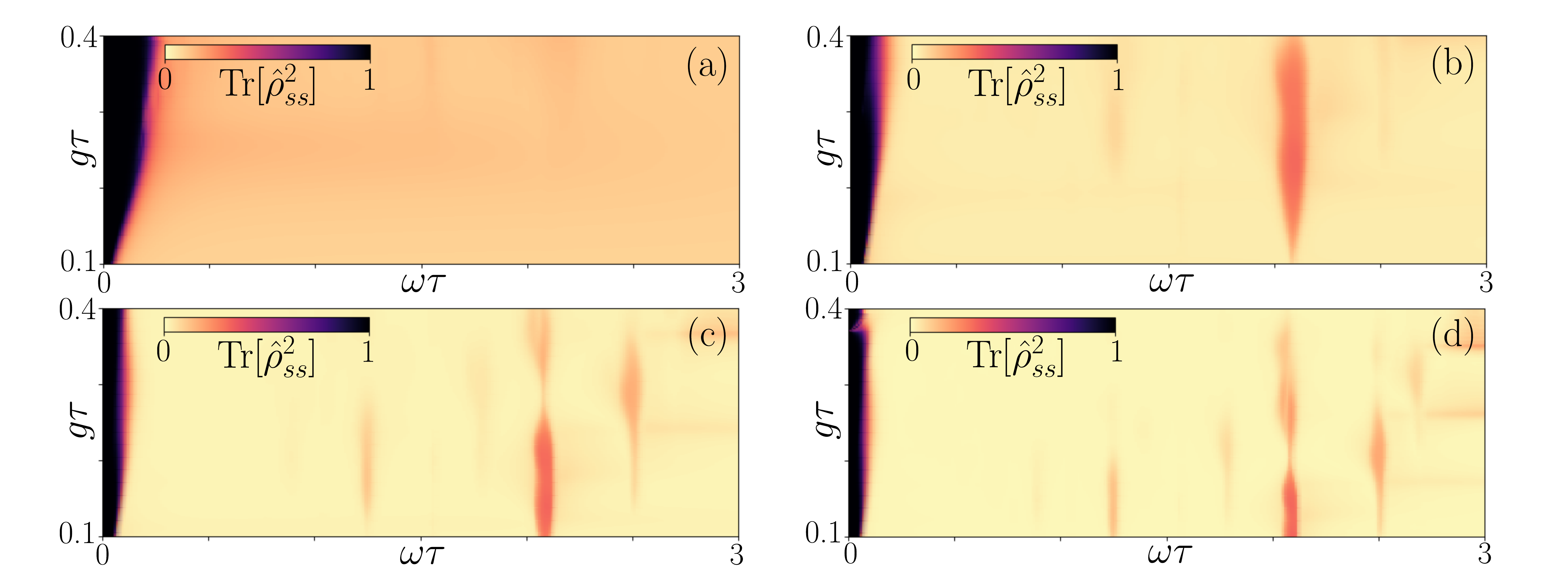}\\
\caption{Stationary purity $\text{Tr}[\hat{\rho}_{\mathrm{ss}}^2]$ varying $g\tau$ and $\omega\tau$ and for different system sizes: (a) $N=10$, (b) $N=30$, (c) $N=50$, (d) $N=70$.}\label{fig_mapssizes}
\end{figure}

Here we analyze the behavior of the purity islands increasing system size. In Fig. \ref{fig_mapssizes} we plot the purity map for: (a) $N=10$; (b) $N=30$; (c) $N=50$; (d) $N=70$. Notice how the background color becomes clearer progressively, as the minimum attainable purity is $1/N$. For $N=10$, we can observe that the region for larger Rabi frequencies is almost uniform, not displaying purity islands. In the rest of the cases, purity islands are clearly visible. We observe that the number of islands and their shape change with system size in a non-trivial way: for larger system sizes more purity islands emerge (compare $N=70$ and $N=50$ with $N=30$), however at the same time these islands seem to shrink with $N$. These results are in accordance with the non-monotonic behavior already observed for the lifetime of the dominant oscillatory mode, and they illustrate how the metastable resonance do not approach smoothly some thermodynamic limit: rather they are a system size resonance.


\begin{thebibliography}{99}
\bibitem{footnote3} We denote the next (different) leading rate as $\gamma_*$. In the $(2,3)$ resonance $\gamma_*=\gamma_3$ as $\lambda_2=\lambda_1^*$.
\end{thebibliography}
\end{document}